\documentclass[letterpaper]{article} 
\usepackage{aaai24}  
\usepackage{times}  
\usepackage{helvet}  
\usepackage{courier}  
\usepackage[hyphens]{url}  
\usepackage{graphicx} 
\usepackage{natbib}  
\usepackage{caption} 
\usepackage{microtype}
\usepackage{algorithm}
\usepackage{algorithmic}
\usepackage{amsfonts,amssymb,amsmath}
\usepackage{booktabs}
\usepackage{pifont}
\usepackage{graphicx}
\usepackage{subfigure}
\usepackage{xcolor}

\usepackage{newfloat}
\usepackage{listings}
\DeclareCaptionStyle{ruled}{labelfont=normalfont,labelsep=colon,strut=off} 
\floatstyle{ruled}
\newfloat{listing}{tb}{lst}{}
\floatname{listing}{Listing}
\newcommand{\algoname}{AdaCCD }

\title{AdaCCD: Adaptive Semantic Contrasts Discovery Based Cross Lingual Adaptation for Code Clone Detection}
\author {
    Yangkai Du\textsuperscript{\rm 1},
    Tengfei Ma\textsuperscript{\rm 2},
    Lingfei Wu\textsuperscript{\rm 3},
    Xuhong Zhang\textsuperscript{\rm 1},
    Shouling Ji\textsuperscript{\rm 1}\footnote{Corresponding Author.}
}

\affiliations {
    \textsuperscript{\rm 1}Zhejiang University\\
    \textsuperscript{\rm 2}Stony Brook University\\
    \textsuperscript{\rm 3}Anytime.AI\\
    \{yangkaidu,zhangxuhong,slj\}@zju.edu.cn, 
    tengfei.ma@stonybrook.edu, 
    lwu@anytime-ai.com,
}

\usepackage{bibentry}

\begin{document}

\maketitle

\begin{abstract}
Code Clone Detection, which aims to retrieve functionally similar programs from large code bases, has been attracting increasing attention. Modern software often involves a diverse range of programming languages. However, current code clone detection methods are generally limited to only a few popular programming languages due to insufficient annotated data as well as their own model design constraints. To address these issues, we present AdaCCD, a novel cross-lingual adaptation method that can detect cloned codes in a new language without annotations in that language. AdaCCD leverages language-agnostic code representations from pre-trained programming language models and propose an Adaptively Refined Contrastive Learning framework to transfer knowledge from resource-rich languages to resource-poor languages. We evaluate the cross-lingual adaptation results of AdaCCD by constructing a multilingual code clone detection benchmark consisting of 5 programming languages. AdaCCD achieves significant improvements over other baselines, and achieve comparable performance to supervised fine-tuning.
\end{abstract}

\section{Introduction}
\newcommand{\ccd}{Code Clone Detection }
\ccd is an essential task in software engineering, which aims to identify functionally similar source code from a large code base. Those identified code snippets benefit programmers in reviewing or refactoring code. In recent years, numerous novel neural models \citep{wei_cdlh_2017,zhang_novel_2019,yu_tbccd_2019,wang_faast_2020} have been proposed for \ccd and achieved great success on popular code clone benchmarks.

However, a major limitation of existing models for code clone detection is that they only support a single programming language. In the real world, large projects often consist of multiple programming languages, including scripting languages like Python and system languages like C/C++ and Rust. This urges us to develop a model that supports multiple languages at the same time.
Additionally, these models often rely on a large amount of annotated data for training, which is not always available for less popular languages. This can limit their ability to detect code clones in projects containing low-resource languages.

One solution is representing code in different languages with a uniform compiler-generated intermediate representation (IR). The model could be trained on IR \cite{ben-nun_neural_2018, VenkataKeerthy_ir2vec_2020} rather than on the source code, allowing it to learn common patterns across different languages. However, obtaining IR for different programming languages requires intensive domain expertise and engineering efforts to fix compilation errors, making it infeasible for language extension. Another solution is utilizing a pre-trained multilingual code encoder. By pre-training programming language models (PPLMs) on large code corpora containing multiple programming languages \cite{feng_codebert_2020,lachaux_dobf_2021,wang-etal-2021-codet5,guo_graphcodebert_2021,guo_unixcoder_2022}, PPLMs can learn language-agnostic code context and representations from self-supervised learning. 
However, the performance using such self-supervised code representations directly for code clone detection is generally much inferior to supervised fine-tuning. To address this limitation, we are exploring the cross-lingual adaptation method, i.e. we first train a PPLM-based model on only one language with annotation data and then adapt it to other languages without annotation data. This could be extremely useful for clone detection on low-resource programming languages.

In this work, we propose a new contrastive learning framework of cross-lingual adaptation for code clone detection named AdaCCD. Given the PPLM-based model pretrained on one language, AdaCCD discovers semantically similar and dissimilar contrasts through clustering and neighborhood search. Considering that the discovered contrastive pairs from clustering are not reliable (especially in the beginning phase), AdaCCD also incorporates semantic-preserving program transformations to derive guaranteed semantic similar contrasts. 
A novel adaptive mechanism is then used to adjust the preference of the two types of contrasts, reducing false positives in the discovery stage. 
Applying contrastive learning to those semantic contrasts encloses the representation of semantically similar programs and pushes away the representation of dissimilar programs in the target language. AdaCCD iteratively improves its accuracy by using the enhanced model to discover more accurate semantic contrasts from unlabeled target corpora and using these for bootstrapping.

In experiments we adapt GraphCodeBERT \cite{guo_graphcodebert_2021} and CodeBERT \cite{feng_codebert_2020}, trained on POJ-104/GCJ benchmarks \cite{Mou_poj104_2016,ye_misim_2020}, to five low-resource languages: Rust, Ruby, JavaScript, Go, and C\# with significant performance improvement. The proposed AdaCCD (with no annotation in target languages) is shown comparable to supervised fine-tuning methods.

Our contributions can be summarized as three folds:
\begin{itemize}
    \item It is the first work focusing on the cross-lingual adaptation for code clone detection, where lack of annotated data for low-resource language is an essential bottleneck.  
    \item We propose AdaCCD, a novel and effective method which leverages language-agnostic representations of PPLMs and contrastive learning for cross-lingual adaptation.
    \item We design a new Adaptively Refined Contrastive Learning framework based on semantic contrasts discovery and adaptive control for balancing different types of contrasts.
\end{itemize}
\section{Related Work}
\subsection{Code Clone Detection}
Code Clone detection aims to extract similar pairs of code snippets from large code bases. A pair of similar codes is called a clone. According to the definition of similar code, code clones can be categorized into four types \cite{roy_ccdreview_18}. We focus on addressing the Type-4 clone, code pairs that are functionally similar, which are the most challenging and cannot be accurately detected by trivial text matching. 

The community has proposed several novel neural models for detecting Type-4 clone \citep{wei_cdlh_2017,zhang_novel_2019,yu_tbccd_2019,wang_faast_2020}, which exploit abstract syntax tree or data flow information obtained from compilers to help understand code functionality. In the training stage, code snippets are encoded into low-dimension dense vectors, and the vector similarity in the latent space is increased for the clone pair, while vector similarity is decreased for the non-clone pair. However, the structure and node property of the abstract syntax tree of each programming language is distinct, so adapting these abstract syntax and data flow-based models to new programming languages is non-trivial.

Unlike these methods, \algoname only uses source code as input. The reason is that we hope our framework can be generalized and easily extended to all programming languages. Additionally, we are the first to focus on the cross-lingual adaptation of the code clone detection task, which can alleviate the lack of annotated data in low-resource languages.

\subsection{Contrastive Learning}
Contrastive learning has been proven to be a promising self-supervised pre-training task for vision \cite{he_moco_2020,chen_simclr_2020,zbontar21a_barlow_21} and language similarity \cite{gao-etal-2021-simcse} tasks. The principle of contrastive learning is to learn invariant representations for different views of an instance generated by data transformation. These different views of the same instance are called positive contrasts, while other instances are called negative contrasts. 
By enclosing the representations of positive contrasts and pushing apart the representations of negative contrasts, the model can learn invariant representations for samples in different views with the same semantics.
There have been a few attempts at using contrastive learning for code similarity learning. These attempts either use a source-to-source compiler to create different views of same program as positive contrasts and use other programs as negative contrasts \cite{jain_contrastive_2021} or use identifier renaming for positive contrast generation and real-world bug injection for negative contrast generation \cite{ding-etal-2022-towards}. 

AdaCCD differs from those methods, which aim to learn uniform representations for different views of the same program while separating representations of different programs. Instead, AdaCCD focuses on enclosing the representation of semantically similar programs while separating representations of dissimilar programs in the target language by our Adaptively Refined Contrastive Learning framework.

\subsection{Multilingual Software Engineering}
\citet{ahmed_multilingual_2022} shed light on the data bottleneck present in low-resource programming languages. They emphasize the similarities found in human-written code across different programming languages and propose that multilingual training for software engineering tasks can greatly benefit these low-resource programming languages. Their findings inspire us to leverage the knowledge acquired from high-resource languages for unseen low-resource languages.

\citet{li2022crosslingual} present PLATO, a cross-lingual adaptation framework aimed at leveraging knowledge acquired from labeled datasets to perform type inference tasks in unlabeled target languages.
\citet{chai2022crossdomain} propose a transfer learning framework, named CDCS, for code-search by adapting a few-shot meta-learning algorithm called MAML \cite{Baik_2021_ICCV} to the code-search task. 
CLCDSA \cite{clcdsa_nafi_2019} propose a cross-language clone detector using syntactical features and API Documentation, but it requires manual efforts for feature extraction and parallel data for finetuning.
In contrast, our work, AdaCCD, is the first to focus on cross-lingual adaptation for code clone detection tasks. We propose a novel Adaptively Refined Contrastive Learning framework specifically tailored for transferring the similarity function learned in labeled source language to unlabeled target language.
\section{Background}
We will first introduce the problem definition of cross-lingual adaptation for the code clone detection task. Then we will present our motivating early findings when conducting zero-shot cross-lingual adaptation using two multilingual PPLMs.
\subsection{Problem Definition}
Given a code snippet $p\in \mathcal{D}$, where $\mathcal{D}$ is a code base containing multiple snippets, the code clone detection task aims to retrieve the set of functionally similar programs of $p$ from $\mathcal{D}$. Under the cross-lingual adaptation setting, an annotated code base $\mathcal{D}_{s}$ in high-resource source language $l_s$ is available, each snippet in $\mathcal{D}_{s}$ is labeled with a set of clones in ${D}_{s}$. However, the code base $\mathcal{D}_{t}$ in low-resource target language $l_t$ is unlabeled. The goal of cross-lingual adaptation is achieving overwhelming code clone performance on $l_t$ with $\mathcal{D}_{s}$ and unlabeled $\mathcal{D}_{t}$. 

\subsection{Code Representation}
\label{sec:code-rep}
We add a two-layer MLP with relu activation after the last layer of the transformer and use the last layer's hidden states of [CLS] token for generating program embedding. Code Clone detection is achieved by cosine similarity measurement between code embeddings.

\subsection{Zero-Shot Cross-Lingual Transfer}
\label{sec:zero-shot}
We start our investigation by first investigating zero-shot cross-lingual transfer performance on CodeBERT, and GraphCodeBERT trained on the POJ-104 \cite{Mou_poj104_2016}, i.e., directly evaluating model performance on 5 programming languages other than C/C++. We use the InfoNCE loss \cite{oord2019representation} as the objective for training on source language. 


\begin{figure}[!ht]
    \centering
    \includegraphics[width=1.0\linewidth]{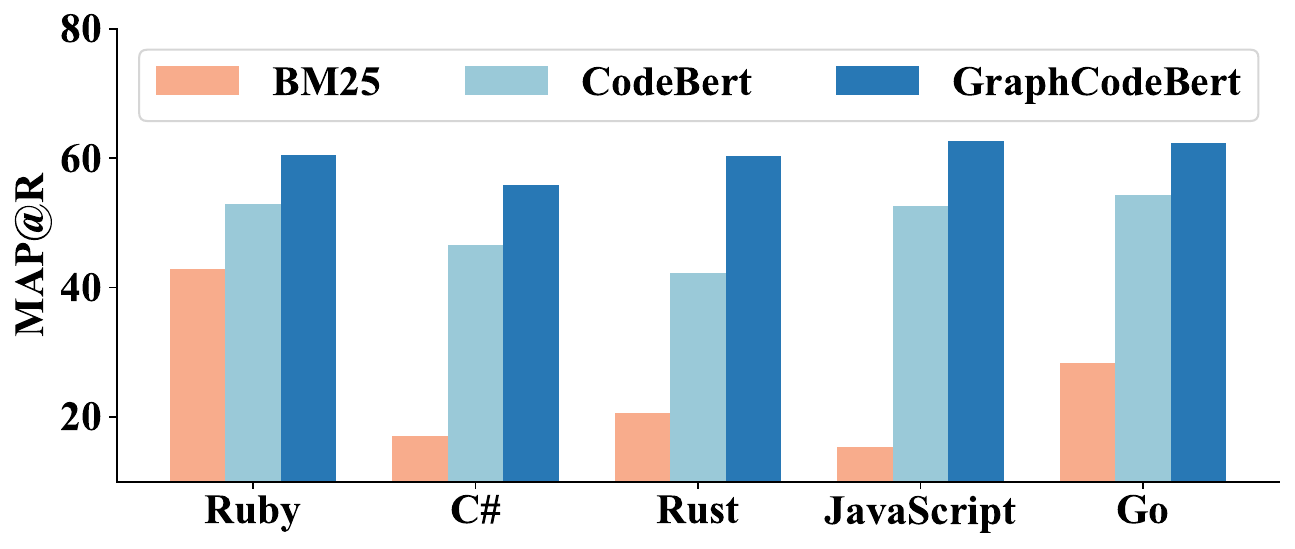}
    \caption{Zero-Shot Adaptation Results. MAP@R: Mean Average Precision @ R, which measures how accurately a model can retrieve similar items given a query.}
    \label{fig:bm25}
\end{figure}

In Figure \ref{fig:bm25}, we compare the zero-shot transfer results of the CodeBERT and GraphCodeBERT models with BM25 \cite{bm25_spark}, a strong baseline for text retrieval. The similarity metric is the pair-wise cosine distance in vector space. Rust and C\# are not among the pretraining languages of PPLMs, so their adaptation performance is relatively lower than the other languages. Nevertheless, both CodeBERT and GraphCodeBERT outperform BM25 in retrieving semantically similar programs from all five programming languages, indicating their ability to generate language-agnostic code representation and promising generalization capabilities for unseen languages. 

The empirical study of PPLM's generalization capabilities to unseen languages motivates us to automatically retrieve semantic similar and dissimilar contrasts from unlabeled corpus on target language using PPLM trained on the source language. 
We can use those discovered contrasts as pseudo-labels for cross-lingual adaptation.
\section{Methods}
To improve the performance of PPLMs on retrieving semantic similar programs from unseen programming languages, we propose a novel cross-lingual adaptation method based on an Adaptively Refined Contrastive Learning framework for Code Clone Detection, i.e., AdaCCD. 

\subsection{Adaptively Refined Contrastive Learning}
\begin{figure}[!ht]
    \centering
    \includegraphics[width=1.0\linewidth]{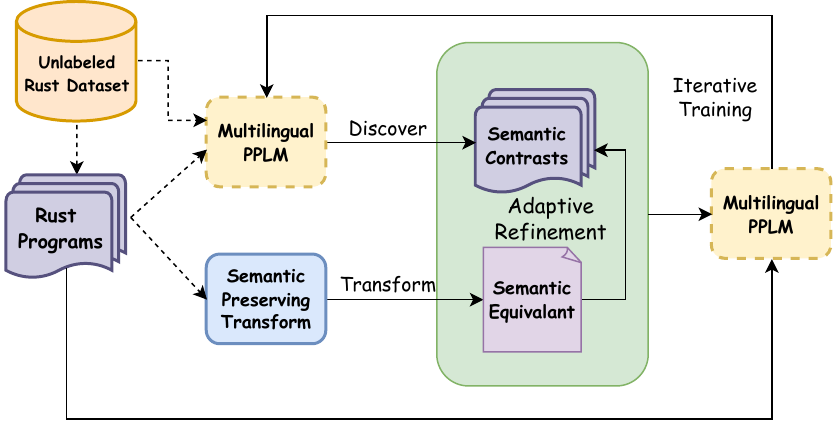}
    \caption{Overview of AdaCCD. We give an example when adapting to Rust language.}
    \label{fig:overview}
\end{figure}

AdaCCD automatically \textbf{discovers semantic similar/dissimilar contrasts} from unlabeled target dataset $\mathcal{D}_t$ by probing the model trained on $\mathcal{D}_s$. 
Furthermore, we introduce an \textbf{adaptive refinement} mechanism that exploits \textbf{semantic-preserving program transformation} and adaptive preference for discovered and transformed contrasts to reduce false positives in the discovered semantic contrasts. Models trained on those refined contrasts can iteratively discover more accurate contrasts, further boosting model performance in the target language. We provide an overview of AdaCCD in Figure \ref{fig:overview}.

\subsection{Adaptive Refinement}
\label{sec:adaptive-comb}
Both the discovered and transformed semantic contrasts 
offer unique benefits and drawbacks. The transformed similar contrasts are derived from the anchor program through semantic-preserving transformations, guaranteeing semantic equivalence without introducing noise to the model. However, these transformed contrasts often exhibit shallow common features due to limitations imposed by transformation methods used, making them easily identifiable. Conversely, the discovered similar contrasts have the advantage of retrieving syntactically diverse programs from various sources, each with distinct code styles. By exposing the model to this complexity, it can learn to recognize more complex semantic features within different code styles, thereby enhancing its capabilities for detecting semantic code clones.

Those two parts can complement each other. We propose an adaptive refinement strategy to dynamically adjust the preference of discovered/transformed similar contrasts throughout the training process. 
For an anchor program $p_i$ at training step $t$, the set of discovered similar contrasts $P$ and a semantic-preserving transformation function, we first randomly sample a discovered similar contrast $p_i^{disc}$ from $P$ and apply the transformation for anchor program $p_i$ to derive a guaranteed similar contrast $p_i^{trans}$. 
Finally we sample the final positive contrast of $p_i$, i.e., $p_i^+$ from $\{p_i^{trans},p_i^{disc}\}$ with $P(p_i^+ = p_i^{trans}) = \alpha_t$.
The higher the $\alpha_t$, the higher the bias towards the transformed contrast. 

We employ a linear decay approach to adjust the preference parameter $\alpha_t$ for discovered/transformed similar contrast throughout the training process. Intuitively, the model struggles to accurately identify semantically similar contrasts during the early stages of training, so the preference for transformed contrasts should be high to reduce noise. However, as the model progressively adapts to unseen language patterns, it becomes beneficial to increase the preference for discovered contrasts. This adjustment enables the introduction of greater complexity and diversity, enhancing the model's capabilities in handling a wider range of semantic variations.
Formally, the $\alpha_t$ is adjusted by Equation \ref{eq:adjust}.
\begin{align}
\label{eq:adjust}
\begin{split}
\alpha_t = \left \{
\begin{array}{lr}
    \alpha_{0},                    & \frac{t}{T_{total}} \leq \sigma \\
    \alpha_{0} \cdot (1 - \frac{t}{T}),     & \frac{t}{T_{total}} > \sigma
\end{array}
\right.
\end{split}
\end{align}
$\alpha_0 \in [0,1]$ is the initial value for $\alpha_t$ at the start of the training,
$T_{total}$ is the total training steps, and $\sigma \in [0,1]$ is a hyperparameter for determining when should preference for transformed contrasts begin to decay. We set $\sigma$ to $0.1$ and provide a detailed discussion for the impact of $\alpha_0$ in our experiment.

\subsection{Semantic Contrasts Discover}
\label{sec:discover}
The zero-shot cross-lingual adaptation results of PPLMs inspire us to automatically discover semantic similar/dissimilar contrasts without supervised signal by probing the model trained on $\mathcal{D}_s$ using unlabeled target dataset $\mathcal{D}_t$.
Prior works have proposed methods for mining parallel texts using multilingual embedding \cite{artetxe-schwenk-2019-margin,schwenk-etal-2021-ccmatrix}, but it is hard to determine a suitable score threshold. Instead, we employ Kmeans clustering for contrasts discover.
Following the aforementioned code embedding process, we first obtain code embedding $\mathbf{x} \in \mathbb{R}^d$ for each program $p$ in $\mathcal{D}_t$. Then we assign each program to one of $C$ semantic classes by running Kmeans clustering on $\{ \mathbf{x}_i \}_{i=1}^N$. We use cosine similarity as the distance metric.
$C$ is a hyperparameter for the number of cluster centers. 
We provide a discussion of the choice of $C$ in our experiment.

\paragraph{Positive Contrasts Discover: }
Because semantically similar programs are more likely to be assigned to the compact space than dissimilar programs, according to our zero-shot study. In order to identify semantically similar programs, the AdaCCD algorithm considers programs assigned to the same cluster as candidates for similar contrasts. The set of candidate programs is denoted as $T$ in the following discussion.
\begin{equation}
    T = \{p_j | p_j \in \mathcal{D}_t  \land c_i = c_j \}
\end{equation}
However, some dissimilar programs would also be assigned unexpectedly. To reduce those false positive cases, we leave out borderline samples from the candidates by only keeping the top $k$ nearest neighborhood of $p_i$. Finally, we formulate the set of discovered semantic similar contrasts $P$ of $p_i$ in equation \ref{eq:discovered-pos-set}.
\begin{equation}
    \label{eq:discovered-pos-set}
    P = \{p_j | p_j \in T  \land rank(s_{ij}) \leq k \}
\end{equation}
Where $s_{ij}$ is the cosine similarity score of $\mathbf{x}_i$ and $\mathbf{x}_j$, and $rank$ is a function for retrieving the sorted index of $s_{ij} \in \{s_{ia} \}_{a=1}^{|T|}$ (in descending order). $k$ is a hyperparameter.

\paragraph{Negative Contrasts Discover: }
One crucial factor for contrastive learning is having diverse negative contrasts, which can prevent the model from collapsing to a trivial solution \cite{gao-etal-2021-simcse}. To meet this requirement, we use all programs not in the same cluster as $p_i$ as the set of negative contrasts, ensuring the model is exposed to sufficient and diverse negative contrasts.
\begin{equation}
    N = \{p_j | p_j \in \mathcal{D}_t  \land c_i \neq c_j \}
\end{equation}
\subsection{Semantic-Preserving Transformation}
While AdaCCD can effectively discover semantic contrasts through clustering and neighborhood selection, it is still possible to include some false positive samples in the discovered set $P$ and can be particularly problematic when the model's performance in the target language is uncompelling. To address this issue, AdaCCD employs semantic-preserving transformation of programs to derive additional positive contrasts, allowing the model to be trained on guaranteed semantic similar contrasts. 
We incorporate two semantic-preserving transformations, Back Translation and Identifier Renaming, which are practical and easily extensible to new programming languages.

\paragraph{Identifier Renaming:}
We use identifier renaming for creating different views of code snippets from all languages. Identifier renaming only alters the identifier names so that it will not change the functionality. For program $p_i$, we randomly apply one of those two renaming strategies: 
\begin{itemize}
    \item Identifier Normalization: normalize the variable names to "var\_1", "var\_2", ..., "var\_m" and function names to "func\_1", "func\_2", ..., "func\_m".
    \item Identifier Randomization: first collect variable and function names from all programs to a name pool, and randomly select names from the pool.
\end{itemize}

\paragraph{Back Translation:}
Back Translation is an effective data augmentation approach in NLP tasks \cite{edunov-etal-2018-understanding}. 
Given the original text in English, it first translates the original text to another language, usually German, by a neural machine translation model, then the German text is translated back to English. The back-translated text is usually in distinct text forms but semantically equivalent when a strong translation model is used. 
Inspired by the success of back translation, we also exploit back translation for program transformation. We use CodeGeeX \cite{zheng2023codegeex}, a multilingual code generation model trained on large-scale code data, to translate programs to Python and then to the original language.

Furthermore, our Adaptively Refined Contrastive Learning framework is not confined to the aforementioned transformations; it can accommodate the integration of additional semantic-preserving transformation functions. 
We provide an in-depth discussion of transformation functions in our experiment.

\subsection{Iterative Training}
\label{sec:iterative}
Updating the discovered contrasts set on every model step is highly time-consuming and even infeasible when $\mathcal{D}_t$ is large since we have to encode all the programs and run Kmeans again to discover new contrasts.
\begin{table*}[!ht]
\centering
\begin{tabular}{p{4.65cm}cccccc}
\toprule
Methods         & Ruby  & C\#   & Rust  & JavaScript & Go & Avg.   \\ 
\midrule
Text-embedding-ada-002  & 71.37 & 47.83 & 59.81 & 50.47 & 59.13 & 54.31 \\
ContraCode-FT        & 65.42 & 59.06 & 63.85 & 53.97      & 67.59 & 61.12\\ 
CodeBERT-FT            & 74.21 & 72.00 & 84.06 & 74.62      & 78.89 & 77.39 \\
GraphCodeBERT-FT       & 80.03 & 74.74 & 87.63 & 76.84      & 82.54 & 80.44 \\
\midrule
CodeBERT-$\text{ZeroTrans}^{\dagger}$    & 51.84 & 44.82 & 49.42 & 51.84  & 53.83 & 49.98 \\
CodeBERT-$\text{Whiten}^{\dagger}$          & 51.18 & 51.03 & 55.92 & 56.34  & 57.00 & 55.07 \\
CodeBERT-$\text{AdaCCD-IR}^{\dagger}$          & \textbf{64.26}/\textbf{65.63} & 53.55/54.77 & 59.46/63.24     & 61.80/62.13  & 61.50/63.88 & 59.08/61.00     \\
CodeBERT-$\text{AdaCCD-BT}^{\dagger}$         & --    & \textbf{60.65}/\textbf{62.19} & \textbf{75.89}/\textbf{76.97}       & \textbf{63.41}/\textbf{62.41}   & \textbf{65.77}/\textbf{71.20}     & \textbf{66.43}/\textbf{68.19}   \\
\midrule
CodeBERT-$\text{ZeroTrans}^{\ddagger}$         & 52.96 & 46.52 & 42.24 & 52.63      & 54.31 & 48.93 \\
CodeBERT-$\text{Whiten}^{\ddagger}$         & 54.05 & 52.12 & 51.11 & 59.89      & 56.66 & 54.95 \\
CodeBERT-$\text{AdaCCD-IR}^{\ddagger}$         & \textbf{57.22}/\textbf{64.05} & 57.74/58.45 & 56.91/59.29 & 64.87/64.52      & 60.90/61.19 & 60.10/60.86 \\
CodeBERT-$\text{AdaCCD-BT}^{\ddagger}$        & --    & \textbf{64.37}/\textbf{65.42} & \textbf{72.86}/\textbf{75.49}  & \textbf{65.74}/\textbf{66.30}      & \textbf{62.55}/\textbf{63.43} & \textbf{66.38}/\textbf{67.66} \\

\midrule
GraphCodeBERT-$\text{ZeroTrans}^{\dagger}$     & 61.98 & 54.86 & 57.46 & 62.95 & 66.65 & 60.48 \\
GraphCodeBERT-$\text{Whiten}^{\dagger}$     & 59.30 & 58.73 & 59.99 & 65.67 & 65.59 & 62.50 \\
GraphCodeBERT-$\text{AdaCCD-IR}^{\dagger}$      & \textbf{71.41}/\textbf{72.35} & 62.55/63.74    & 74.44/74.65 & \textbf{72.86}/\textbf{73.58} & 72.41/76.08  & 70.57/72.01 \\
GraphCodeBERT-$\text{AdaCCD-BT}^{\dagger}$      & --    & \textbf{64.96}/\textbf{66.95} & \textbf{79.20}/\textbf{79.22} & 70.82/72.38 & \textbf{72.82}/\textbf{77.56}  & \textbf{71.95}/\textbf{74.03}   \\

\midrule
GraphCodeBERT-$\text{ZeroTrans}^{\ddagger}$     & 60.47 & 55.84 & 60.29 & 62.60      & 62.34 & 60.27\\
GraphCodeBERT-$\text{Whiten}^{\ddagger}$    & 58.75 & 59.55 & 63.87 & 66.07      & 64.74 & 63.56\\
GraphCodeBERT-$\text{AdaCCD-IR}^{\ddagger}$   & \textbf{68.80}/72.40 & 64.40/64.42 & 73.50/74.34 & 68.84/69.65      & 70.13/71.88 & 69.22/70.06 \\
GraphCodeBERT-$\text{AdaCCD-BT}^{\ddagger}$ & --    & \textbf{67.28}/\textbf{69.82} & \textbf{82.35}/\textbf{83.23} & \textbf{70.56}/\textbf{71.31} & \textbf{69.74}/\textbf{72.77} &  \textbf{72.48}/\textbf{74.28} \\
\toprule

\end{tabular}
\caption{Main results. We report two test MAP@R for AdaCCD; using default hyper setting/using validation data for hyperparameter search.
$\dagger$ denote transfered from GCJ and $\ddagger$ denote transfered from POJ-104. The CodeGeex model currently do not support translation for Ruby, so we cannot implement AdaCCD-BT on Ruby benchmark. Avg. is the average MAP@R of C\#, Rust, JavaScript and Go benchmarks. All results are reported with 3 random seeds.}
\label{tab:main_results}
\end{table*}

To reduce the training time cost, we only update the discovered contrastive sets of each program at the start of epochs. Furthermore, we randomly split $\mathcal{D}_t$ into two parts, $\mathcal{D}_t^1$ and $\mathcal{D}_t^2$, to reduce the pool size for clustering and neighborhood search, which need pair-wise distance measurement. We adopted the training strategy of alternating between $\mathcal{D}_t^1$ and $\mathcal{D}_t^2$. At the start of the training, we first discover contrastive sets on $\mathcal{D}_t^1$ and train the model on $\mathcal{D}_t^1$. Then we alternate $\mathcal{D}_t^1$ to $\mathcal{D}_t^2$, use the enhanced model to discover more accurate contrastive sets on $\mathcal{D}_t^2$, and repeat.

At the adaptive training stage, we randomly sample a batch of programs $\{p_1,p_2,\dots ,p_b \}$ from $\mathcal{D}_t^1$ or $\mathcal{D}_t^2$, each program $p_i$ is associated with a sampled positive contrast $p_i^+$, positive contrasts set $P_i$ and negative contrasts set $N_i$. 
We concatenate $\{p_1^+,p_2^+,\dots ,p_b^+ \}$ to $\{p_1,p_2,\dots ,p_b \}$, and feed those 2B programs to the model. Finally, We use the contrastive loss $\mathcal{L}_{ad}$ for adaptive training:

\begin{equation}
    - \sum_{p_i} \sum_{p_j \in P_i} log\left( \frac{e^{s_{ij}/\tau}}{e^{s_{ij}/\tau} + \sum_{p_k \in N_i} e^{s_{ik}/\tau}}\right)
    \label{eq:ft-loss}
\end{equation}
Note that to make equations neat, we make a few changes to $P_i$ and $N_i$. The $p_i^+$ is added to $P_i$, and the $p_j^+$ is added to $N_i$ if $p_j \in N_i$. $\tau$ is a temperature hyperparameter.

\section{Experiment}
To validate that our \algoname can generalize to different multilingual PPLMs and source datasets, We use CodeBERT and GraphCodeBERT models for all experiments. Besides, We train CodeBERT and GraphCodeBERT on two popular C/C++ datasets, POJ-104 and GCJ. 
\subsection{Evaluation Benchmark \& Setting}
\label{sec:exp-setting}
To evaluate the cross-lingual adaptation performance of \algoname on unseen languages, we collect 5 datasets in Rust, Ruby, JavaScript, Go, and C\# language from CodeNet \cite{puri_codenet_21}. We randomly split each dataset to train/validation/test splits. Validation data is used for performance tuning of supervised finetuning methods and reporting the optimal performance of AdaCCD when labeled validation data is provided. 
If labeled data is not available in real-world use case, 
$k$ is set to $16$ uniformly across all languages, and $\alpha_0$ is set to $0.8$ except for Go ($0.4$ instead).
We use the Mean Average Precision @ R (MAP@R) \cite{musgrave_mapr_2020} as the evaluation metric following \citet{ye_misim_2020}. Our code and data will be released for future research.

\subsection{Cross Lingual Adaptation Results}
\paragraph{Comparison to Baselines\\}
We first investigate whether \algoname can effectively improve model performance on unseen programming languages with only unlabeled data on the target language. To answer this research question, we compare our \algoname with the following baseline settings.
\textbf{Text-embedding-ada-002:} One of the most powerful text embedding API services provided by OpenAI \cite{openaiemb2022}. It can encode the text and code to a dense vector and tackle text and code similarity tasks without finetuning, including code-search-code task, which is similar to code clone detection.
\textbf{ZeroTrans:} This setting is zero-shot cross-lingual transfer as introduced in the Background section.
\textbf{Whiten:} BERT-Whitening \cite{su_whitenbert_21} is a simple but effective approach for improving unsupervised text semantic similarity matching on target dataset by transforming co-variance of sentence representation to an identity matrix.
\textbf{FT:} Directly finetune the model using labeled data of the target language. We investigate two \algoname variations using Identifier Renaming and Back Translation, denoted as AdaCCD-IR and AdaCCD-BT. The results are listed in Table \ref{tab:main_results}. 
The proposed \algoname can significantly improve the cross-lingual adaptation performance in those 5 languages without using labeled data compared to unsupervised baselines. The improvement can be generalized to models trained on different source data. 
In addition, when a small number of valiadation data is provided for performance tuning, \algoname is comparable to finetuning with labeled data. \algoname on GraphCodeBERT outperforms finetuning ContraCode and is comparable to finetuning CodeBERT.

\paragraph{Back Translation Vs. Identifier Renaming\\}
Both Identifier Renaming and Back Translation variations are effective on CodeBERT and GraphCodeBERT Models. According to our observations, using Back Translation for program transformation is superior to Identifier Renaming. We attribute the superiority of Back Translation to producing more syntactically diverse positive samples, which encourage model to learn program similarity from semantic features rather than shallow syntactical features. Nevertheless, Identifier Renaming is still a simple and effective choice.

\subsection{Ablation Study}
In order to reduce the false discovered semantic contrasts, we propose to refine the discovered contrasts with transformed contrasts derived from semantic-preserving program transformation. We conduct an ablation study to validate whether those two proposed components positively contribute to adaptation results. We list results in Table \ref{tab:ablation}. \ding{56} \textit{adapt.} represent the setting that we remove the adaptive control of the preference for transformed and discovered contrasts and replace it with a constant preference $\alpha_0$ throughout the training process. \ding{56} \textit{trans.} denote that we do not incorporate transformed contrasts and only use discovered semantic contrasts for training. We implement it by fixing $\alpha _t$ to $0$ throughout training. For all the settings, we report the test MAP@R using the hyperparameters as introduced in Section \ref{sec:exp-setting}.
\begin{table}[!ht]
\centering
\begin{tabular}{lcccc}
\toprule
Setting                 & C\#   & Rust  & JS & Go \\
\midrule
\textsc{CB-IR}              & \textbf{57.74} & \textbf{56.91} & \textbf{64.87} & \textbf{60.90} \\
\ding{56} \textit{adapt.}  & 57.66 & 53.11 & 63.71 & 60.49 \\     
\ding{56} \textit{trans.} & 56.89 & 54.77 & 62.72 & 59.35 \\
\midrule
\textsc{CB-BT}              & \textbf{64.37} & 72.86  & \textbf{65.74}      & \textbf{62.55} \\
\ding{56} \textit{adapt.}  & 62.38 & \textbf{74.78} & 64.06 & 57.87 \\     
\ding{56} \textit{trans.}    & 56.89 & 54.77 & 62.72 & 59.35 \\

\midrule
\textsc{GCB-IR}              & 64.40 & 73.50 & 68.84      & \textbf{70.13} \\
\ding{56} \textit{adapt.}  & 63.42 & 70.37 & 67.75 & 68.43 \\     
\ding{56} \textit{trans.} & \textbf{64.53} & \textbf{75.47} & \textbf{70.11} & 69.43 \\
\midrule
\textsc{GCB-BT}              &\textbf{67.28} & \textbf{82.35} & \textbf{70.56} & \textbf{69.74} \\
\ding{56} \textit{adapt.}  & 64.71 & 81.09 & 67.28 & 63.77 \\     
\ding{56} \textit{trans.}    & 64.53 & 75.47 & 70.11 & 69.43 \\

\toprule
\end{tabular}
\caption{Ablation Study. We conduct ablation study by adapting from POJ-104 dataset. CB/GCB denote using CodeBERT/GraphCodeBERT as backbone. 
}
\label{tab:ablation}
\end{table}

Under CB-IR, CB-BT, and GCB-BT settings, the adaptation performance drops when we cancel transformed contrasts or adaptive control of preference. However, reducing those two components can occasionally benefit the performance under the GCB-IR setting, where GraphCodeBERT is the backbone model and uses Identifier Renaming for transformation. 
We attribute the reason for this to the difference in pretraining tasks of GraphCodeBERT and CodeBERT.
Compared with CodeBERT, GraphCodeBERT exploits the def-use graph of the variables in the pretraining stage. 
They incorporate variable sequences from the def-use graph and design two identifier-aware tasks, data-flow edge prediction and variable-code alignment. 
Therefore, the contrasts transformed by Identifier Renaming cannot provide new information for semantic similarity learning under the GraphCodeBERT setting.

\subsection{Model Analysis}
\begin{figure}[!htb]
  \centering
    \subfigure[C\#]{\includegraphics[width=0.46\linewidth]{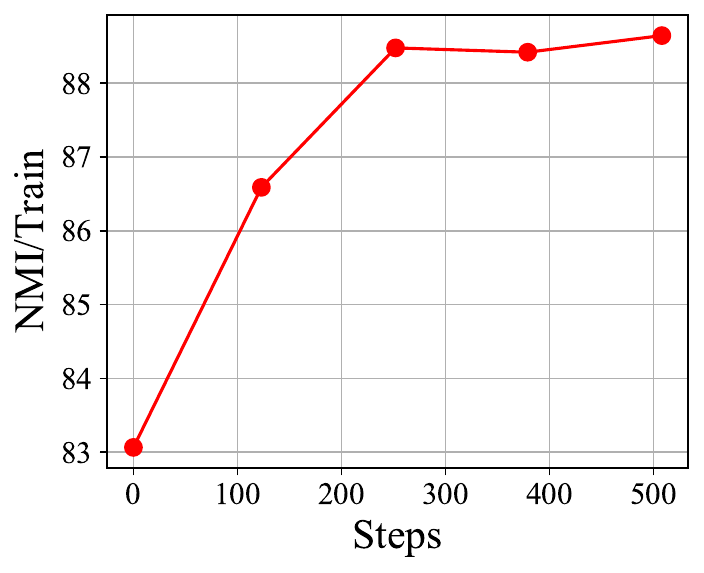}} \hspace{0.0\linewidth}
	\subfigure[Rust]{\includegraphics[width=0.46\linewidth]{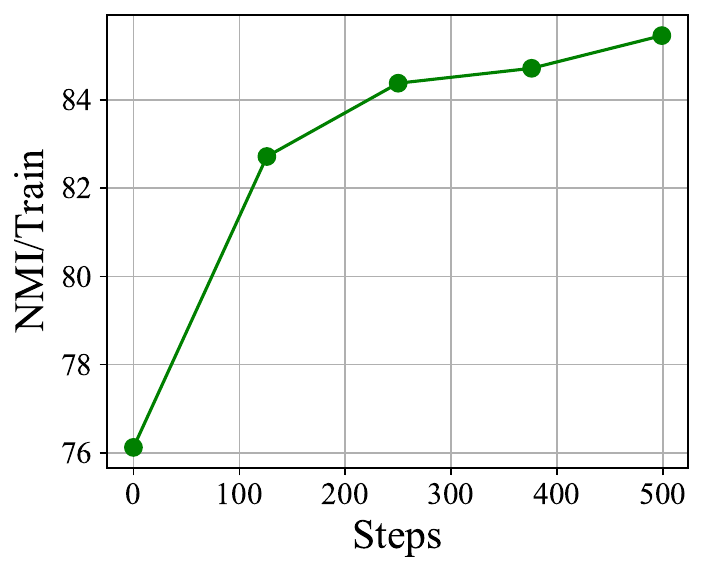}} \vfill
    \subfigure[JavaScript]{\includegraphics[width=0.46\linewidth]{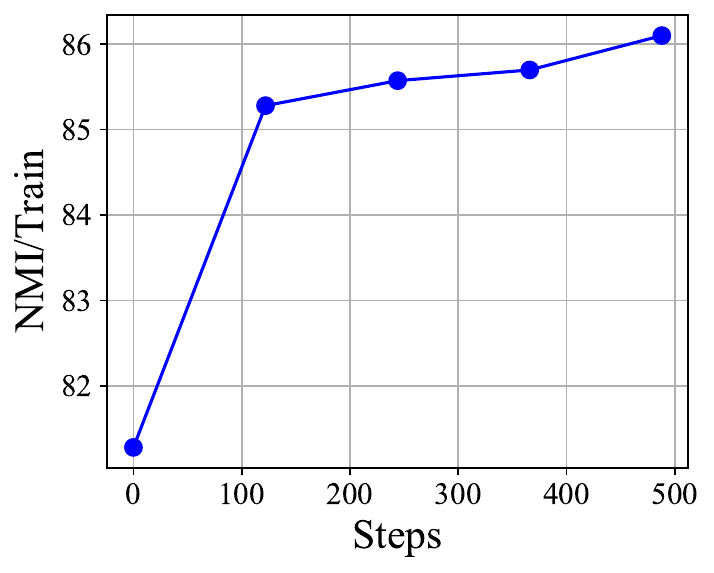}} \hspace{0.0\linewidth}
	\subfigure[Go]{\includegraphics[width=0.46\linewidth]{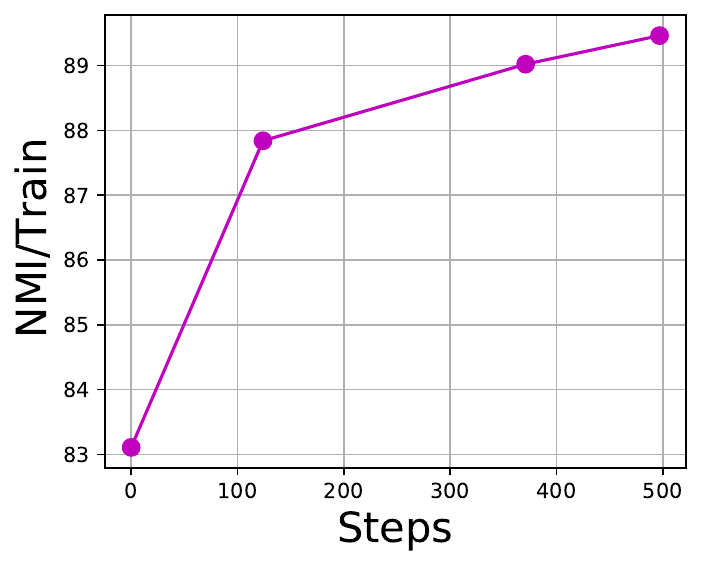}}
  \caption{NMI evaluated at the end of epochs. }
	\label{fig:nmi}
	\vspace{0.0in}
\end{figure}
\paragraph{Discover Quality: } Our method is an iterative framework. The model can only be bootstrapped when discovered contrasts' accuracy improves. The discovered contrasts are determined by Kmeans clustering, so the accuracy of discovered contrasts can be measured by Kmeans clustering quality. 
We use Normalized Mutual Information (NMI) between clustering ids and ground-truth problems ids for clustering quality measurement following the practice in deep clustering methods \cite{Caron_2018_ECCV}. 
NMI is measured under the GCB-BT setting, using the POJ-104 as the source dataset.

On all 4 target languages, the NMI gradually increases throughout the training, indicating that the match between the clustering-assigned IDs and the ground-truth problem IDs becomes closer over time. Therefore, we can conclude that the accuracy of the discovered contrasts improves during the training process, allowing the model to be successfully bootstrapped under our iterative framework.
\paragraph{Choice of $\alpha_0$: }
\begin{figure}[!htb]
  \centering
    \subfigure[C\#]{\includegraphics[width=0.46\linewidth]{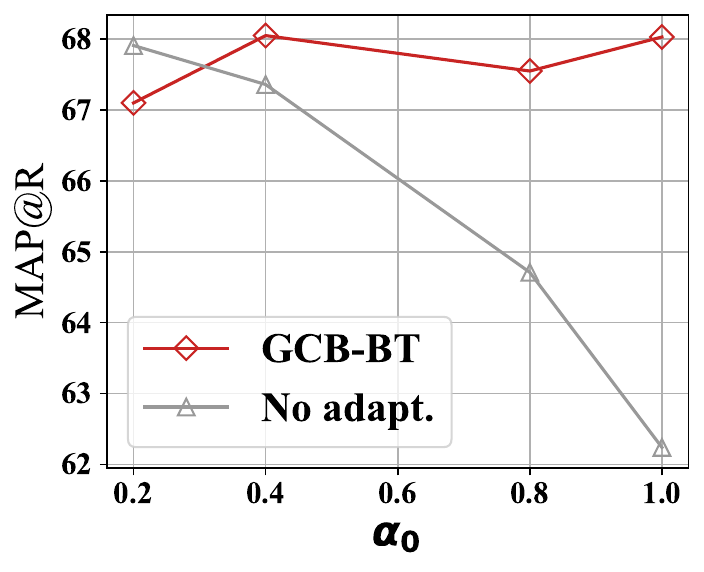}} \hspace{0.0\linewidth}
	\subfigure[Rust]{\includegraphics[width=0.46\linewidth]{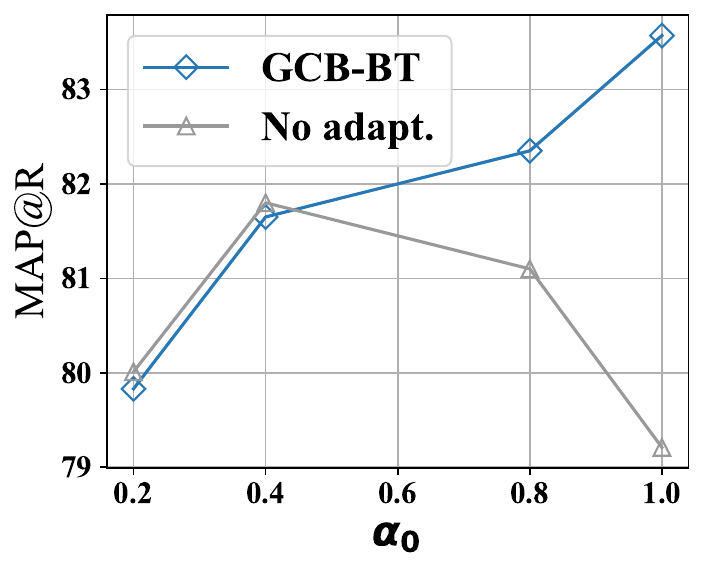}} \vfill
    \subfigure[JavaScript]{\includegraphics[width=0.46\linewidth]{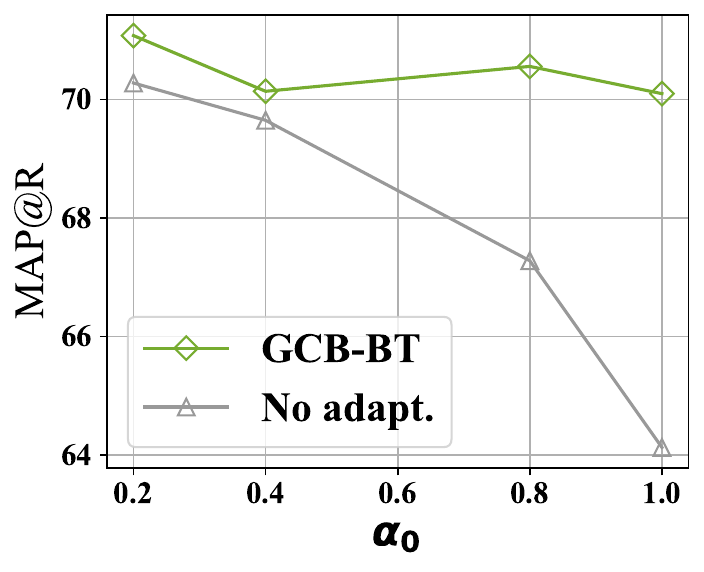}} \hspace{0.0\linewidth}
	\subfigure[Go]{\includegraphics[width=0.46\linewidth]{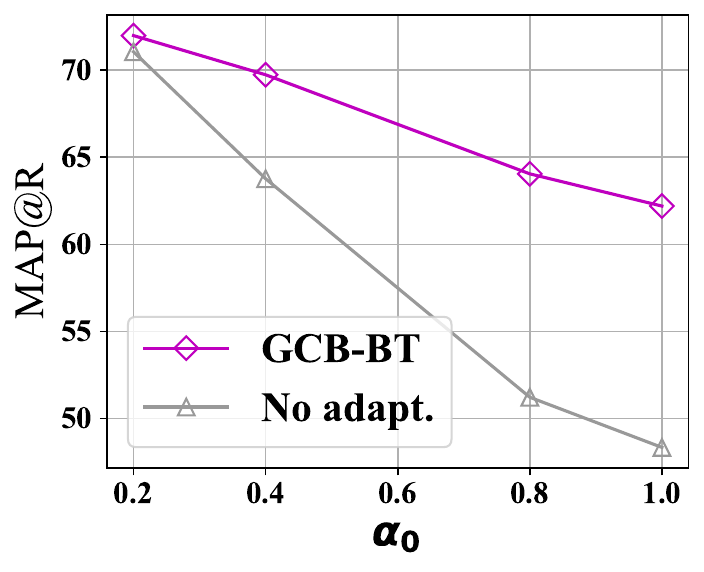}}
  \caption{Sensitivity test of $\alpha_0$ under GCB-BT setting. We conduct sensitivity test by adapting from POJ-104.}
	\label{fig:sensitivity}
	\vspace{0.0in}
\end{figure}
$\alpha_0$ is an important hyper-parameter for the adaptive control of preference for transformed and discovered contrasts. The higher $\alpha_0$, the transformed contrasts participate more throughout the training process. To test the model sensitivity of $\alpha_0$, we plot the model performance against $\alpha_0$ under the GCB-BT setting in Figure \ref{fig:sensitivity}. Besides, we also plot \ding{56} \textit{adapt.} setting introduced in the ablation study for comparison. We observe that \algoname is more robust to choices of $\alpha_0$ compared with the ablation setting, where a constant value replaces adaptive control for $\alpha$. The worst performance of our methods can at least beat the zero-shot adaptation, which cannot be guaranteed when removing adaptive control for $\alpha_0$, e.g., $\alpha_0 = 0.8$ or $1.0$ on the Go benchmark. Moreover, \algoname is superior to the ablation setting under most cases.

\paragraph{Choice of $C$: }
We set the cluster number $C$ for Kmeans to the number of programming problems in $\mathcal{D}_{t}$. This setting is based on the assumption that the number of functionalities in $\mathcal{D}_{t}$ is known. It is not always realistic, especially when we conduct adaptation in the wild. Therefore, we investigate whether $C$ should be strictly set to the number of functionalities in $\mathcal{D}_{t}$ to ensure the adaptation performance of AdaCCD. Specifically, we halve/double the original $C$ and conduct adaptation under the GCB-BT setting. The results are reported in Table \ref{tab:choice_c}. 

\begin{table}[!ht]
\centering
\begin{tabular}{lcccc}
\toprule
Setting                 & C\#   & Rust  & JS & Go \\
\midrule
\textsc{GCB-BT}              & \textbf{67.28} & \textbf{82.35} & \textbf{70.56} & 69.74 \\
\textit{Half}  & 67.79 & 81.95 & 70.22 & 70.51 \\     
\textit{Double}    & 68.13 & 82.82 & 70.38 & \textbf{71.80} \\

\toprule
\end{tabular}
\caption{\textit{Half}/\textit{Double} represents that $C$ is set to half/double of problem numbers.}
\label{tab:choice_c}
\end{table}
We observe that the performance of \textit{Half} slightly degrade, and surprisingly, the performance of \textit{Double} even outperforms original setting.
These results reveal that AdaCCD is robust to choices of $C$ when selected from an acceptable range and still has space for improvement when more resource is given for hyperparameter search. Overall, AdaCCD is still effective even if the problem number is unavailable, and the choice of $C$ is not strictly constrained.
\paragraph{Choice of Program Transformations: }
In order to assess the influence of program transformations and showcase the versatility of our AdaCCD, beyond Back-Translation and Identifier Renaming, we conduct experiments involving an additional transformation technique, i.e., Code Rewriting. This involves rewriting the program utilizing the exceptional code generation capabilities of GPT-3.5-turbo \cite{openai2022gpt35}. The results are presented in Table \ref{tab:variant}. Our observations indicate that AdaCCD is not limited by specific program transformations, as it can effectively accommodate any valid transformations. 
\begin{table}[!ht]
\centering
\begin{tabular}{lcccc}
\toprule
Setting              & Rust  & JS & Go & Avg. \\
\midrule

\textsc{GCB-IR}      & 73.50 & 68.84 & 70.13 & 70.82 \\
\textsc{GCB-BT}      & \textbf{82.35} & 70.56 & 69.74 & 74.22 \\
\textsc{GCB-CR}      & 75.33 & \textbf{71.91} & \textbf{79.47} & \textbf{75.57} \\

\toprule
\end{tabular}
\caption{CR denote using code rewriting as transformations. We conduct experiment by adapting from POJ-104.}
\label{tab:variant}
\end{table}
\section{Conclusion}
In this paper, we present AdaCCD, a novel cross-lingual adaptation method for code clone detection based on adaptive semantic contrasts discovery. 
Our experiments show that AdaCCD can achieve significant performance improvements on unlabeled languages and achieve comparable performance with supervised finetuning methods. 
Our work enabling models to analyze code clone in diverse languages, where the lack of annotated data for lesser-used languages is a significant bottleneck.

\clearpage
\section*{Acknowledgements}
This work was partly supported by NSFC under No. U1936215 and 62102360.
\bibliography{aaai24}

\begin{thebibliography}{37}
\providecommand{\natexlab}[1]{#1}

\bibitem[{Ahmed and Devanbu(2022)}]{ahmed_multilingual_2022}
Ahmed, T.; and Devanbu, P. 2022.
\newblock Multilingual training for {Software} {Engineering}.
\newblock \emph{arXiv:2112.02043 [cs]}.
\newblock ArXiv: 2112.02043.

\bibitem[{Artetxe and Schwenk(2019)}]{artetxe-schwenk-2019-margin}
Artetxe, M.; and Schwenk, H. 2019.
\newblock Margin-based Parallel Corpus Mining with Multilingual Sentence Embeddings.
\newblock In Korhonen, A.; Traum, D.; and M{\`a}rquez, L., eds., \emph{Proceedings of the 57th Annual Meeting of the Association for Computational Linguistics}, 3197--3203. Florence, Italy: Association for Computational Linguistics.

\bibitem[{Baik et~al.(2021)Baik, Choi, Kim, Cho, Min, and Lee}]{Baik_2021_ICCV}
Baik, S.; Choi, J.; Kim, H.; Cho, D.; Min, J.; and Lee, K.~M. 2021.
\newblock Meta-Learning With Task-Adaptive Loss Function for Few-Shot Learning.
\newblock In \emph{Proceedings of the IEEE/CVF International Conference on Computer Vision (ICCV)}, 9465--9474.

\bibitem[{Ben-Nun, Jakobovits, and Hoefler(2018)}]{ben-nun_neural_2018}
Ben-Nun, T.; Jakobovits, A.~S.; and Hoefler, T. 2018.
\newblock Neural code comprehension: a learnable representation of code semantics.
\newblock In \emph{Proceedings of the 32nd {International} {Conference} on {Neural} {Information} {Processing} {Systems}}, {NIPS}'18, 3589--3601. Red Hook, NY, USA: Curran Associates Inc.

\bibitem[{Caron et~al.(2018)Caron, Bojanowski, Joulin, and Douze}]{Caron_2018_ECCV}
Caron, M.; Bojanowski, P.; Joulin, A.; and Douze, M. 2018.
\newblock Deep Clustering for Unsupervised Learning of Visual Features.
\newblock In \emph{Proceedings of the European Conference on Computer Vision (ECCV)}.

\bibitem[{Chai et~al.(2022)Chai, Zhang, Shen, and Gu}]{chai2022crossdomain}
Chai, Y.; Zhang, H.; Shen, B.; and Gu, X. 2022.
\newblock Cross-Domain Deep Code Search with Meta Learning.
\newblock In \emph{Proceedings of the 44th International Conference on Software Engineering}, ICSE '22, 487–498. New York, NY, USA: Association for Computing Machinery.
\newblock ISBN 9781450392211.

\bibitem[{Chen et~al.(2020)Chen, Kornblith, Norouzi, and Hinton}]{chen_simclr_2020}
Chen, T.; Kornblith, S.; Norouzi, M.; and Hinton, G. 2020.
\newblock A Simple Framework for Contrastive Learning of Visual Representations.
\newblock In III, H.~D.; and Singh, A., eds., \emph{Proceedings of the 37th International Conference on Machine Learning}, volume 119 of \emph{Proceedings of Machine Learning Research}, 1597--1607. PMLR.

\bibitem[{Ding et~al.(2022)Ding, Buratti, Pujar, Morari, Ray, and Chakraborty}]{ding-etal-2022-towards}
Ding, Y.; Buratti, L.; Pujar, S.; Morari, A.; Ray, B.; and Chakraborty, S. 2022.
\newblock Towards Learning (Dis)-Similarity of Source Code from Program Contrasts.
\newblock In \emph{Proceedings of the 60th Annual Meeting of the Association for Computational Linguistics (Volume 1: Long Papers)}, 6300--6312. Dublin, Ireland: Association for Computational Linguistics.

\bibitem[{Edunov et~al.(2018)Edunov, Ott, Auli, and Grangier}]{edunov-etal-2018-understanding}
Edunov, S.; Ott, M.; Auli, M.; and Grangier, D. 2018.
\newblock Understanding Back-Translation at Scale.
\newblock In \emph{Proceedings of the 2018 Conference on Empirical Methods in Natural Language Processing}, 489--500. Brussels, Belgium: Association for Computational Linguistics.

\bibitem[{Feng et~al.(2020)Feng, Guo, Tang, Duan, Feng, Gong, Shou, Qin, Liu, Jiang, and Zhou}]{feng_codebert_2020}
Feng, Z.; Guo, D.; Tang, D.; Duan, N.; Feng, X.; Gong, M.; Shou, L.; Qin, B.; Liu, T.; Jiang, D.; and Zhou, M. 2020.
\newblock {CodeBERT}: {A} {Pre}-{Trained} {Model} for {Programming} and {Natural} {Languages}.
\newblock \emph{arXiv:2002.08155 [cs]}.
\newblock ArXiv: 2002.08155.

\bibitem[{Gao, Yao, and Chen(2021)}]{gao-etal-2021-simcse}
Gao, T.; Yao, X.; and Chen, D. 2021.
\newblock {S}im{CSE}: Simple Contrastive Learning of Sentence Embeddings.
\newblock In \emph{Proceedings of the 2021 Conference on Empirical Methods in Natural Language Processing}, 6894--6910. Online and Punta Cana, Dominican Republic: Association for Computational Linguistics.

\bibitem[{Guo et~al.(2022)Guo, Lu, Duan, Wang, Zhou, and Yin}]{guo_unixcoder_2022}
Guo, D.; Lu, S.; Duan, N.; Wang, Y.; Zhou, M.; and Yin, J. 2022.
\newblock {UniXcoder}: {Unified} {Cross}-{Modal} {Pre}-training for {Code} {Representation}.
\newblock Number: arXiv:2203.03850 arXiv:2203.03850 [cs].

\bibitem[{Guo et~al.(2021)Guo, Ren, Lu, Feng, Tang, Liu, Zhou, Duan, Svyatkovskiy, Fu, Tufano, Deng, Clement, Drain, Sundaresan, Yin, Jiang, and Zhou}]{guo_graphcodebert_2021}
Guo, D.; Ren, S.; Lu, S.; Feng, Z.; Tang, D.; Liu, S.; Zhou, L.; Duan, N.; Svyatkovskiy, A.; Fu, S.; Tufano, M.; Deng, S.~K.; Clement, C.; Drain, D.; Sundaresan, N.; Yin, J.; Jiang, D.; and Zhou, M. 2021.
\newblock {GraphCodeBERT}: {Pre}-training {Code} {Representations} with {Data} {Flow}.
\newblock \emph{arXiv:2009.08366 [cs]}.
\newblock ArXiv: 2009.08366.

\bibitem[{He et~al.(2020)He, Fan, Wu, Xie, and Girshick}]{he_moco_2020}
He, K.; Fan, H.; Wu, Y.; Xie, S.; and Girshick, R. 2020.
\newblock Momentum Contrast for Unsupervised Visual Representation Learning.
\newblock In \emph{Proceedings of the IEEE/CVF Conference on Computer Vision and Pattern Recognition (CVPR)}.

\bibitem[{Jain et~al.(2021)Jain, Jain, Zhang, Abbeel, Gonzalez, and Stoica}]{jain_contrastive_2021}
Jain, P.; Jain, A.; Zhang, T.; Abbeel, P.; Gonzalez, J.; and Stoica, I. 2021.
\newblock Contrastive {Code} {Representation} {Learning}.
\newblock In \emph{Proceedings of the 2021 {Conference} on {Empirical} {Methods} in {Natural} {Language} {Processing}}, 5954--5971. Online and Punta Cana, Dominican Republic: Association for Computational Linguistics.

\bibitem[{Lachaux et~al.(2021)Lachaux, Roziere, Szafraniec, and Lample}]{lachaux_dobf_2021}
Lachaux, M.-A.; Roziere, B.; Szafraniec, M.; and Lample, G. 2021.
\newblock DOBF: A Deobfuscation Pre-Training Objective for Programming Languages.
\newblock In Ranzato, M.; Beygelzimer, A.; Dauphin, Y.; Liang, P.; and Vaughan, J.~W., eds., \emph{Advances in Neural Information Processing Systems}, volume~34, 14967--14979. Curran Associates, Inc.

\bibitem[{Li et~al.(2022)Li, Xie, Li, Xu, Li, and Liu}]{li2022crosslingual}
Li, Z.; Xie, X.; Li, H.; Xu, Z.; Li, Y.; and Liu, Y. 2022.
\newblock Cross-Lingual Transfer Learning for Statistical Type Inference.
\newblock arXiv:2107.00157.

\bibitem[{Mou et~al.(2016)Mou, Li, Zhang, Wang, and Jin}]{Mou_poj104_2016}
Mou, L.; Li, G.; Zhang, L.; Wang, T.; and Jin, Z. 2016.
\newblock Convolutional neural networks over tree structures for programming language processing.
\newblock In \emph{Proceedings of the Thirtieth AAAI Conference on Artificial Intelligence}, 1287--1293.

\bibitem[{Musgrave, Belongie, and Lim(2020)}]{musgrave_mapr_2020}
Musgrave, K.; Belongie, S.; and Lim, S.-N. 2020.
\newblock A Metric Learning Reality Check.

\bibitem[{Nafi et~al.(2019)Nafi, Kar, Roy, Roy, and Schneider}]{clcdsa_nafi_2019}
Nafi, K.~W.; Kar, T.~S.; Roy, B.; Roy, C.~K.; and Schneider, K.~A. 2019.
\newblock CLCDSA: Cross Language Code Clone Detection using Syntactical Features and API Documentation.
\newblock In \emph{2019 34th IEEE/ACM International Conference on Automated Software Engineering (ASE)}, 1026--1037.

\bibitem[{OpenAI(2022{\natexlab{a}})}]{openai2022gpt35}
OpenAI. 2022{\natexlab{a}}.
\newblock Introducing ChatGPT.
\newblock \url{https://openai.com/blog/chatgpt}.
\newblock Accessed: 2022-11-30.

\bibitem[{OpenAI(2022{\natexlab{b}})}]{openaiemb2022}
OpenAI. 2022{\natexlab{b}}.
\newblock New and improved embedding model.
\newblock \url{https://openai.com/blog/new-and-improved-embedding-model}.
\newblock Accessed: 2022-12-15.

\bibitem[{Puri et~al.(2021)Puri, Kung, Janssen, Zhang, Domeniconi, Zolotov, Dolby, Chen, Choudhury, Decker, Thost, Thost, Buratti, Pujar, Ramji, Finkler, Malaika, and Reiss}]{puri_codenet_21}
Puri, R.; Kung, D.; Janssen, G.; Zhang, W.; Domeniconi, G.; Zolotov, V.; Dolby, J.~T.; Chen, J.; Choudhury, M.; Decker, L.; Thost, V.; Thost, V.; Buratti, L.; Pujar, S.; Ramji, S.; Finkler, U.; Malaika, S.; and Reiss, F. 2021.
\newblock CodeNet: A Large-Scale AI for Code Dataset for Learning a Diversity of Coding Tasks.
\newblock In Vanschoren, J.; and Yeung, S., eds., \emph{Proceedings of the Neural Information Processing Systems Track on Datasets and Benchmarks}, volume~1.

\bibitem[{Roy and Cordy(2018)}]{roy_ccdreview_18}
Roy, C.~K.; and Cordy, J.~R. 2018.
\newblock Benchmarks for software clone detection: A ten-year retrospective.
\newblock In \emph{2018 IEEE 25th International Conference on Software Analysis, Evolution and Reengineering (SANER)}, 26--37.

\bibitem[{Schwenk et~al.(2021)Schwenk, Wenzek, Edunov, Grave, Joulin, and Fan}]{schwenk-etal-2021-ccmatrix}
Schwenk, H.; Wenzek, G.; Edunov, S.; Grave, E.; Joulin, A.; and Fan, A. 2021.
\newblock {CCM}atrix: Mining Billions of High-Quality Parallel Sentences on the Web.
\newblock In Zong, C.; Xia, F.; Li, W.; and Navigli, R., eds., \emph{Proceedings of the 59th Annual Meeting of the Association for Computational Linguistics and the 11th International Joint Conference on Natural Language Processing (Volume 1: Long Papers)}, 6490--6500. Online: Association for Computational Linguistics.

\bibitem[{{Sparck Jones}, Walker, and Robertson(2000)}]{bm25_spark}
{Sparck Jones}, K.; Walker, S.; and Robertson, S. 2000.
\newblock A probabilistic model of information retrieval: development and comparative experiments: Part 1.
\newblock \emph{Information Processing \& Management}, 36(6): 779--808.

\bibitem[{Su et~al.(2021)Su, Cao, Liu, and Ou}]{su_whitenbert_21}
Su, J.; Cao, J.; Liu, W.; and Ou, Y. 2021.
\newblock Whitening Sentence Representations for Better Semantics and Faster Retrieval.
\newblock \emph{CoRR}, abs/2103.15316.

\bibitem[{van~den Oord, Li, and Vinyals(2019)}]{oord2019representation}
van~den Oord, A.; Li, Y.; and Vinyals, O. 2019.
\newblock Representation Learning with Contrastive Predictive Coding.
\newblock arXiv:1807.03748.

\bibitem[{VenkataKeerthy et~al.(2020)VenkataKeerthy, Aggarwal, Jain, Desarkar, Upadrasta, and Srikant}]{VenkataKeerthy_ir2vec_2020}
VenkataKeerthy, S.; Aggarwal, R.; Jain, S.; Desarkar, M.~S.; Upadrasta, R.; and Srikant, Y.~N. 2020.
\newblock IR2VEC: LLVM IR Based Scalable Program Embeddings.
\newblock \emph{ACM Trans. Archit. Code Optim.}, 17(4).

\bibitem[{Wang et~al.(2020)Wang, Li, Ma, Xia, and Jin}]{wang_faast_2020}
Wang, W.; Li, G.; Ma, B.; Xia, X.; and Jin, Z. 2020.
\newblock Detecting {Code} {Clones} with {Graph} {Neural} {Networkand} {Flow}-{Augmented} {Abstract} {Syntax} {Tree}.
\newblock \emph{arXiv:2002.08653 [cs]}.
\newblock ArXiv: 2002.08653.

\bibitem[{Wang et~al.(2021)Wang, Wang, Joty, and Hoi}]{wang-etal-2021-codet5}
Wang, Y.; Wang, W.; Joty, S.; and Hoi, S.~C. 2021.
\newblock {C}ode{T}5: Identifier-aware Unified Pre-trained Encoder-Decoder Models for Code Understanding and Generation.
\newblock In \emph{Proceedings of the 2021 Conference on Empirical Methods in Natural Language Processing}, 8696--8708. Online and Punta Cana, Dominican Republic: Association for Computational Linguistics.

\bibitem[{Wei and Li(2017)}]{wei_cdlh_2017}
Wei, H.-H.; and Li, M. 2017.
\newblock Supervised Deep Features for Software Functional Clone Detection by Exploiting Lexical and Syntactical Information in Source Code.
\newblock In \emph{Proceedings of the 26th International Joint Conference on Artificial Intelligence}, IJCAI'17, 3034–3040. AAAI Press.
\newblock ISBN 9780999241103.

\bibitem[{Ye et~al.(2020)Ye, Zhou, Venkat, Marcus, Tatbul, Tithi, Hasabnis, Petersen, Mattson, Kraska, Dubey, Sarkar, and Gottschlich}]{ye_misim_2020}
Ye, F.; Zhou, S.; Venkat, A.; Marcus, R.; Tatbul, N.; Tithi, J.~J.; Hasabnis, N.; Petersen, P.; Mattson, T.; Kraska, T.; Dubey, P.; Sarkar, V.; and Gottschlich, J. 2020.
\newblock MISIM: A Neural Code Semantics Similarity System Using the Context-Aware Semantics Structure.

\bibitem[{Yu et~al.(2019)Yu, Lam, Chen, Li, Xie, and Wang}]{yu_tbccd_2019}
Yu, H.; Lam, W.; Chen, L.; Li, G.; Xie, T.; and Wang, Q. 2019.
\newblock Neural {Detection} of {Semantic} {Code} {Clones} {Via} {Tree}-{Based} {Convolution}.
\newblock In \emph{2019 {IEEE}/{ACM} 27th {International} {Conference} on {Program} {Comprehension} ({ICPC})}, 70--80.
\newblock ISSN: 2643-7171.

\bibitem[{Zbontar et~al.(2021)Zbontar, Jing, Misra, LeCun, and Deny}]{zbontar21a_barlow_21}
Zbontar, J.; Jing, L.; Misra, I.; LeCun, Y.; and Deny, S. 2021.
\newblock Barlow Twins: Self-Supervised Learning via Redundancy Reduction.
\newblock In Meila, M.; and Zhang, T., eds., \emph{Proceedings of the 38th International Conference on Machine Learning}, volume 139 of \emph{Proceedings of Machine Learning Research}, 12310--12320. PMLR.

\bibitem[{Zhang et~al.(2019)Zhang, Wang, Zhang, Sun, Wang, and Liu}]{zhang_novel_2019}
Zhang, J.; Wang, X.; Zhang, H.; Sun, H.; Wang, K.; and Liu, X. 2019.
\newblock A {Novel} {Neural} {Source} {Code} {Representation} {Based} on {Abstract} {Syntax} {Tree}.
\newblock In \emph{2019 {IEEE}/{ACM} 41st {International} {Conference} on {Software} {Engineering} ({ICSE})}, 783--794.
\newblock ISSN: 1558-1225.

\bibitem[{Zheng et~al.(2023)Zheng, Xia, Zou, Dong, Wang, Xue, Wang, Shen, Wang, Li, Su, Yang, and Tang}]{zheng2023codegeex}
Zheng, Q.; Xia, X.; Zou, X.; Dong, Y.; Wang, S.; Xue, Y.; Wang, Z.; Shen, L.; Wang, A.; Li, Y.; Su, T.; Yang, Z.; and Tang, J. 2023.
\newblock CodeGeeX: A Pre-Trained Model for Code Generation with Multilingual Evaluations on HumanEval-X.
\newblock arXiv:2303.17568.

\end{thebibliography}

\end{document}